\documentclass[12pt]{article}
\usepackage{amsmath,amssymb,amsthm,amsfonts,url}

\newcommand{\beq}{\begin{equation}}
\newcommand{\eeq}{\end{equation}}
\newcommand{\bsplit}{\begin{split}}
\newcommand{\esplit}{\end{split}}

\newcommand{\A}{\text{A}}
\newcommand{\B}{\text{B}}

\newtheorem{defi}{Definition}

\newtheorem*{stata}{Statement \A}
\newtheorem*{statb}{Statement \B}

\begin{document}

%%%%%%%%%%%%%%%%%%%%%%%%%%%%%%%%%%%%%%%%%%%%%%%%%%%%%%%%%%%%%%%%%%%%%%%%%%%%%%%%%%%%%%%%%%

\title{Can quantum-mechanical description of physical reality
be considered \emph{in}complete?}

\author{
Gilles Brassard \hspace{0.75cm} Andr\'e Allan M\'ethot\\[0.5cm]
\normalsize\sl D\'epartement d'informatique et de recherche op\'erationnelle\\[-0.1cm]
\normalsize\sl Universit\'e de Montr\'eal, C.P.~6128, Succ.\ Centre-Ville\\[-0.1cm]
\normalsize\sl Montr\'eal (QC), H3C 3J7~~\textsc{Canada}\\ 
\normalsize\texttt{\{brassard,\,methotan\}}\textsf{@}\texttt{iro.umontreal.ca}
}

\date{13 June 2005}

\maketitle

\begin{abstract}
In loving memory of 
Asher Peres, we discuss a
most important and influential paper written in 1935 by his thesis
supervisor and mentor Nathan Rosen, together with
Albert Einstein and Boris Podolsky.
In~that paper, the trio known as EPR 
questioned the completeness of quantum mechanics. The authors argued
that the then-new theory should not be considered final because
they believed it incapable of describing physical reality.
The~epic battle between Einstein and Bohr intensified
following the latter's response later the same year.
Three decades elapsed before John~S.~Bell gave a devastating
proof that the EPR argument was fatally flawed.
The modest purpose of our paper is to give a critical analysis
of the original EPR paper and point out its logical shortcomings
in a way that could have been done 70~years ago, with no need
to wait for Bell's theorem.
We also present an overview of Bohr's response in the interest
of showing how \emph{it} failed to address the gist of the EPR argument.
\end{abstract}

\begin{flushright}
\textit{Dedicated to the memory of Asher Peres}
\end{flushright}

\thispagestyle{empty}

\newpage

%%%%%%%%%%%%%%%%%%%%%%%%%%%%%%%%%%%%%%%%%%%%%%%%%%%%%%%%%%%%%%%%%%%%%%%%%%%%%%%%%%%%%%%%%%

\section{Introduction}\label{sec:intro}

In 1935, Albert Einstein, Boris Podolsky and Nathan Rosen published a paper
that sent shock waves in the physics community~\cite{epr35}, especially in Copenhagen.
L\'eon Rosenfeld wrote about the impact this paper had on Niels Bohr,
one of the fathers of quantum mechanics and its hardiest
defender at the time, that
``this onslaught came down to us as a bolt from the blue''~\cite{bolt}.
Quantum mechanics (QM) was still a relatively young theory, which had 
arisen from several experimental facts.
It~exhibited so many weird features that Bohr once said:
``Anyone who is not shocked by quantum theory has not understood~it''\,%
\footnote{\,Or something of the kind: we can find a variety of similar ``quotes''
attributed to Bohr and it's not clear what he really said!
See~\url{http://en.wikiquote.org/wiki/Niels_Bohr}.}.
Among the non-classical effects of QM, we note its fundamentally
probabilistic nature and the fact that we cannot determine simultaneously
the position and momentum of a particle with arbitrary precision.
In~fact, forget about \textit{measuring} them: QM~tells us that these two conjugate variables
cannot have simultaneous \textit{existence}. These exotic features did not
appeal to everyone in the physics community. To~some, they were an indication that
QM was incapable of fully describing physical reality.
So, with this thought in mind,
Einstein, Podolsky and Rosen (EPR) published the celebrated paper entitled
``Can~quantum-mechanical description of physical reality be
considered complete?''~\cite{epr35}, in which they put forth an argu\-men\-tation
purportedly answering their own question in the negative.
Seventy years after its publication, this is \emph{still} the most heavily
downloaded paper from the American Physical Society website
\url{www.aps.org}\,!

Even though there is evidence~\cite{chris,jammer} that Einstein had been thinking about
it as early as 1931, the notion of \emph{entanglement} was first published in
the 1935 EPR paper. The authors used the correlations obtained from bipartite
measurements of an entan\-gled state to claim that it is possible for the
position and momentum of a particle to have simultaneous existence.
Since the formalism of QM precludes such a possibility,
they concluded that QM cannot offer a complete description of physical reality.
Even though the term ``local hidden variable'' (LHV) was yet
to be coined, it follows in their opinion that nature has to feature
LHVs that cannot be known collectively through any experiment---for~otherwise
the Heisenberg uncertainty relations would be violated---but~never\-theless
these variables determine the behaviour of the physical system
under measurement. Einstein then spent much of the rest of
his life in the vain search of such a theory.

Almost thirty years elapsed before a paper by John~S.
Bell doomed EPR's quest for a LHV theory of nature~\cite{bell64}.
More specifically, Bell defined the most general LHV model possible
and showed that no such model can yield the correlations
predicted by QM when the entangled system proposed by EPR---or~more
precisely its variation subsequently proposed by
David Bohm~\cite{bohm51}---is~measured,
proving in effect that the EPR argument could \emph{not} show that QM is
incomplete. Nevertheless, the situation could have been worse
for quantum mechanics: it~could have been outright incorrect!
Subsequent experi\-ments~\cite{clauser,aspect} established 
the supremacy of QM over LHV theories,
a~somewhat ironic backfiring of the intent of the EPR paper.
It~is a great pity that Einstein did not live to see these breakthroughs.

Even though it took the better part of three decades before Bell
was able to reject all possible LHV models, EPR's conclusions
were debated right away. Two months after the publication of the
EPR paper, Bohr submitted a response
to the same
journal, with the exact same title~\cite{bohr35}.
Quite obviously, Einstein was left entirely unconvinced by Bohr's response.
More on this in Section~\ref{bohr}.

We wish to present here, in loving memory of Asher Peres, a
hopefully insightful \mbox{discussion} of the EPR argument.
The originality of our approach, from a contemporary perspective, is that
its purpose is to poke holes in the the EPR paper \emph{without any
reference to Bell's \mbox{inequality}}. Consequently, its subtitle could have been:
``Why~we would have questioned EPR independently of Bohr's response,
even before Bell came along''!
The~holes we poke are nowhere as fatal as Bell's and they do \emph{not}
rule out the possibility of a LHV theory, which was therefore still a legitimate
field of research until the fateful year~1964.
Instead, we show that the EPR exposition of their ideas
needed polishing for it is marred by several \emph{logical} errors.

After this Introduction, we summarize in Section~\ref{sec:epr}
the EPR argument by quoting the
most relevant excerpts from their paper.
Section~\ref{logic} is the main contribution of our paper:
we~rephrase those quotes in the language
of mathematical logic, which allows us to analyse precisely the
meaning of the words used by~EPR, and exhibit logical holes
in them.
We~address the issue of \emph{spukhafte Fernwirkungen} in
Section~\ref{epistemic}.
In~Section~\ref{bohr}, we present an overview of Bohr's
response to the EPR paper in the interest
of showing how it failed to address the gist of the EPR argument.
Finally, we conclude in Section~\ref{sec:conclusion}.

%%%%%%%%%%%%%%%%%%%%%%%%%%%%%%%%%%%%%%%%%%%%%%%%%%%%%%%%%%%%%%%%%%%%%%%%%%%%%%%%%%%%%%%%%%

\section{The Einstein-Podolsky-Rosen Argument}\label{sec:epr}

As for what should be considered a valid theory of nature, EPR
proposed the following two criteria: (1)~the theory should be
correct and (2)~the theory should be complete. Of~course, the
scientific paradigm should also demand falsifiability as a criterion of
validity, but the EPR argument hinges upon the \emph{completeness}
(or~purported lack of completeness) of~QM\@.
Specifically, the authors give the following definition.\footnote{\,%
Throughout this paper, we set quotes in \textsf{sans serif} whenever
they are taken verbatim from the EPR paper~\cite{epr35} or Bohr's response~\cite{bohr35},
and we use (sans serif) italics in strict accordance with the quoted material.}

\begin{defi}[Completeness]\label{def:completeness}
\textsf{\textit{Every element of the physical reality must have a counterpart in
the physical theory.}}
\end{defi}

The definition of a \emph{physical reality} given by EPR, which
they ``\textsf{regard as reasonable}'' with no further apparent need for justification,
is spelt out in a very famous quote from their paper.

\begin{defi}[Physical reality]\label{def:reality}
\textsf{\textit{If, without in any way disturbing a system, we can predict with
certainty (i.e., with probability equal to unity) the value of a
physical quantity, then there exists an element of physical
reality corresponding to this physical quantity.}}
\end{defi}

As we shall see, the EPR argument is centred around two
statements.

\begin{stata}
Quantum mechanical description of reality is not complete.
\end{stata}

\begin{statb}
Non-commuting operators cannot have simultaneous reality.
\end{statb}

They use both of these statements in an intricate manner in order
to come up with the final ``conclusion'' that Statement~{\A} must be true.
The~structure of their proof can be established from the
following two quotes, which appear at the end of the first section of their paper.

\begin{itemize}
\item[(a)]
\textsf{either (1) \emph{the quantum-mechanical
description of reality  given by the wave function is not
complete} or (2) \emph{when operators corresponding to two
physical quantities do not commute the two quantities cannot have
simultaneous reality}.}

\item[(b)]
\textsf{In quantum mechanics it is usually assumed that the wave function
\emph{does} contain a complete description of the physical reality
of the system in the state to which it corresponds.
[\ldots]
We~shall show, however, that this assumption, together with the
criterion of reality given above, leads to a contradiction.}

\end{itemize}

\noindent
The argument culminates in their conclusion:

\begin{itemize}
\item[(c)]
\textsf{Starting then with the assumption that the wave function does give
a complete description of the physical reality, we arrived at the
conclusion that two physical quantities, with noncommuting
operators, can have simultaneous reality.  [\ldots]
We~are thus forced to
conclude that the quantum-mechanical description of physical
reality given by wave functions is not complete.}
\end{itemize}

\section{The Language of Logic}\label{logic}

To extract the substance of the EPR argument, we shall now rephrase its structure
in the \mbox{language} of mathematical logic. In~this language, Quote~(a) translates to:
``Either Statement~{\A} or Statement~{\B}''.
The normal usage of ``\emph{either}~$p$ or $q$'' in
English corresponds mathematically to the \emph{exclusive-or} of $p$ and $q$,
namely that either $p$ or $q$ is true \emph{but not both}.
This is written \mbox{$p \oplus q$} in modern mathematical notation.
Therefore, EPR's first quote amounts to saying that
\begin{equation}\label{eq:claima}
\A \oplus \B \, .
\end{equation}

Quote~(b) is a little more subtle to translate into the language
of mathematical logic. First of all, we must understand that
``\textsf{the~criterion of reality given above}'', at the end of the quote,
corresponds to Statement~{\B} being true.
Therefore, Quote~(b) translates to
\mbox{$\neg \A \wedge \B \Rightarrow \textsf{false}$}.
This is mathematically equivalent to saying that 
\begin{equation}\label{eq:claimb}
\neg \A \Rightarrow \neg \B
\end{equation}
as can be verified easily using a truth table.
Proposition~(\ref{eq:claimb}) is stated more explicitly
in the first sentence of Quote~(c) above.

\newpage

Given that EPR eventually derive their desired conclusion,
namely Statement~{\A}, \mbox{after} ``estab\-lishing'' the truth
of both quotes, their final argument---Quote~(c)---hinges upon logical \mbox{tautology}
\mbox{$(\ref{eq:claima}) \wedge (\ref{eq:claimb}) \Rightarrow \A$}:
\begin{equation}\label{eq:tautology}
(\A \oplus \B) \wedge (\neg \A \Rightarrow \neg \B) \Rightarrow \A \, .
\end{equation}

Having set the stage for our logical analysis of the EPR argument,
we now proceed to discussing these three propositions one by~one.

\subsection{The First Proposition}\label{sc:first}

In order to establish $\A \oplus \B$, it must be shown that either
Statement~{\A} or Statement~{\B} holds. To~do this, EPR
used the fact that the formalism of QM does not allow for
a particle's position and momentum to be simultaneously defined.
Therefore, if position and momentum do indeed have
simultaneous reality (negation of Statement~{\B}) then QM
cannot be complete (Statement~{\A}). In~other words,
\mbox{$\neg \B \Rightarrow \A$}. Furthermore, they also
argued that  if QM is already a complete theory
(negation of Statement~{\A}) then it must be the case
that position and momentum cannot have
simultaneous reality (Statement~{\B}). In~other words,
\mbox{$\neg \A \Rightarrow \B$}.

At this point, we remark that working through
both implications was redundant since they are logically equivalent,
each one being the contrapositive of the other.
A~more serious problem with the EPR argumentation is that the
desired conclusion~(1) does \emph{not} follow!
Indeed, \mbox{$\neg \A \Rightarrow \B$} is logically equivalent
to \mbox{$\A \vee \B$}, which cannot be used to conclude
that \mbox{$\A \oplus \B$}.
Worse: no amount of logical reasoning can save the day
since nowhere do EPR argue that it's not possible for
both Statements~{\A} and~{\B} to hold, and indeed that would
be possible should QM be incomplete for some other reason!

We suspect that this problem with the EPR argument simply comes from
our (linguistically correct) interpretation of the word ``either''
in Quote~(a) from Section~\ref{sec:epr}: Instead of Proposition~(\ref{eq:claima}),
no doubt EPR had
\begin{equation}\label{eq:claimabis}
\A \vee \B
\end{equation}
in mind, which indeed is correct, even by today's account.
This happened \emph{either} because EPR were not careful in their use
of the English language, or because the meaning of ``either'' has evolved
in the past seventy years.
But~there is no reason to worry further about this issue since we shall see
in Section~\ref{sc:tautology} that it has no real impact on the EPR conclusion.

\subsection{The Second Proposition}

In order to ``prove'' $\neg \A \Rightarrow \neg \B$,
Einstein, Podolsky and Rosen
used a bipartite entangled state of two particles, where two physicists,
say Alice and Bob, take one particle each and move to space-like
separated regions. Let~us assume that Alice performs a
momentum measurement on her particle. From the measurement
postulate of QM, she is then able to predict with certainty the
outcome of a momentum measurement on Bob's particle. Since Alice and
Bob are in space-like separated regions, there cannot be any
further interaction between the particles once they are separated.
Therefore, Alice's measurement cannot have disturbed Bob's system
in any way. Hence, according to Definition~\ref{def:reality},
momentum corresponds to an element of physical reality for Bob's particle.
Let us change the
assumption to a position measurement on Alice's part of the
entangled state, \emph{instead of} a momentum measurement.
Once again, she will be able to predict the result of a
position measurement on Bob's particle without having disturbed~it.
Therefore, position also corre\-sponds to an element of
physical reality for Bob's particle.
It~``follows'' that position and momentum (two non-commuting operators) had to
have simultaneous reality before either measurement.
This ``establishes'' that \B\ is~false.

The flaw, of course, is that
EPR's argument hinged upon counterfactual arguments of the kind
``the~physicist could do this \emph{or} he could do that'' in cases when
clearly he could not do both.
Nobody put this flaw in better light than Asher Peres, in his
famous aphorism:
``Unperformed experiments have no results''~\cite{noresults}.
Since Alice cannot perform both measurements, it is not the case
that both outcomes are defined simultaneously.
Being fully aware that Einstein would have been left utterly unconvinced
by this reasoning, we take here a different line of attack,
based on logic rather than physics.

Recall that we were supposed to prove~$\neg \A \Rightarrow \neg \B$.
The careful reader will have noticed that EPR reached their desired~``$\neg \B$''
conclusion \emph{without having at any point invoked the assumption}~``$\neg \A$''\@!
Indeed, the reasoning did \emph{not} hinge upon the completeness of QM or lack thereof.
Instead, it hinged upon the \emph{correctness} of~QM, in particular the correctness
of the measurement postulate. Therefore, if we are willing to assume that QM is
correct in terms of its predictions about what can be observed---which we believed
to have been EPR's point of~view---then the EPR argument that purports to prove
\mbox{$\neg \A \Rightarrow \neg \B$} in fact directly ``proves''~$\neg \B$.

\subsection{The Tautology}\label{sc:tautology}

{From} a purely logical point of view, there is nothing to complain about here\ldots\ at~least
on first approach.  Proposition~(\ref{eq:tautology}) is indeed a tautology, as is easily seen
by use of a truth table or a rather simple logical argument\,%
\footnote{\,Assume~\A\ to be false. Then~\B\ is true since $\A \oplus \B$.
But then \A\ would be true since \mbox{$\neg\A\ \Rightarrow \neg\B$} is equivalent to \mbox{$\B \Rightarrow \A$} and \B\ is true.
This contradicts the assumption that \A\ is false, and it must therefore be the case that \A\ is true.}.
However, this tautology is not genuinely relevant to the EPR argument.
First of all, as discussed in Section~\ref{sc:first}, Proposition~(\ref{eq:claima}) should be
replaced by Proposition~(\ref{eq:claimabis}) in the tautology~(\ref{eq:tautology}), yielding
\begin{equation}\label{eq:tautologybis}
(\A \vee \B) \wedge (\neg \A \Rightarrow \neg \B) \Rightarrow \A \, .
\end{equation}
The good news is that Proposition~(\ref{eq:tautologybis}) is also a tautology,
whence the issue of the precise meaning of ``either\ldots or'' turns out to
be moot, as promised at the end of Section~\ref{sc:first}.

But now remember that, notwithstanding Quote~(b), the reasoning given by EPR does \emph{not} prove Proposition~(\ref{eq:claimb})
since they used nowhere the assump\-tion ``\mbox{$\neg\A$}'' that QM is complete. Rather, they ``proved'' \mbox{$\neg\B$}
directly, under the sole assumption that QM is \emph{correct}---in~particular the measurement postulate---which
seems to be taken for granted throughout the paper so strongly that it's not worth calling it an ``assumption''.
Therefore, the somewhat complicated-looking tautology that we wrote down as Proposition~(\ref{eq:tautology}),
to translate the English written words of~EPR, boils down in the end to nothing more than:
\begin{equation}\label{eq:triviality}
(\A \vee \B) \wedge \neg \B \Rightarrow \A \, .
\end{equation}
This statement is such a triviality\,%
\footnote{\,If~{\A} or {\B} is true and it is not {\B}, then it must be~{\A}.
Please compare this with the reasoning spelt out in the previous footnote!}
that we are forced to conclude that the original tautology was a smoke-screen at best.

In summary, the EPR conclusion may have been correct had they been allowed to make counterfactual reasoning,
but their logic to get there was unnecessarily intricate and even outright wrong in some places.

\newpage
\section{Spukhafte Fernwirkungen}\label{epistemic}

Consider the following quote from the EPR paper, which corresponds to the situation after
an interaction has left two particles in the singlet state. 

\begin{quote}\sf
We can then calculate with the help of Schr\"odinger's equation the state
of the combined system [\ldots].
We~cannot, however, calculate the state in which either one of the
two systems is left after the interaction. This, according to
quantum mechanics, can be done only with the help of further
measurements, by a process know as the \emph{reduction of the wave
packet}.
\end{quote}

\noindent
{From} a modern perspective, this is utter nonsense!
\emph{Of~course} we can calculate the state of either system by use of
partial tracing: it's the completely mixed state.
Obviously, EPR considered that only \emph{pure} states deserve being
called ``states''.
Given their insistence on ``elements of reality'', this position
is hardly surprising.
Nevertheless, we must point out that von Neumann had already introduced
density matrices~\cite{vonneumann32}
a few years before EPR wrote their~paper.

If we take the epistemic viewpoint according to which the density matrix
corresponding to Bob's possible knowledge is all there is to the state of his particle,
then Alice's measurement in a space-like separated region has no instantaneous
influence on that state.
It's only if and when Alice tells Bob the result of her measurement that
the state of Bob's particle changes \emph{at~Bob's}, but this communication cannot
take place faster than the speed of light: \emph{spukhafte Fernwirkungen} no more!\,%
\footnote{\,With apologies to Spider-Man! \texttt{:-)}}

We mentioned the above in the interest of readers who may still be under the delusion
that Alice's measurement has an instantaneous influence on the state of Bob's particle.
However, we acknowledge that this issue is not central to a discussion of the EPR paper
since they took the unambiguous (and correct) position that such influence would not take place.
Their mistake was in taking the ontic position that all states are pure, from which it follows that
the state of Bob's particle is defined independently of Alice's or Bob's knowledge about it.

\section{Bohr's Response}\label{bohr}

In his response, Bohr~\cite{bohr35} hardly addresses the issue of entanglement.
For him, the phenomenon described in the EPR paper is nothing more
than what he calls \emph{complementarity}. He~claims that in~QM, as well
as in general relativity, we must take into account the measuring
device to make predictions on the object of interest. In~essence,
Bohr argues that position and momentum cannot both be measured
with an arbitrary precision
\emph{because we cannot do so experimentally}, even in principle.
A~measure of the momentum of a particle can only
be experimentally realized by a momentum transfer on the
measurement apparatus, therefore creating a displacement and
forsaking forever the knowledge of the position of the particle.
In~the EPR argument, says Bohr, when Alice measures the momentum of her
particle, she learns that of Bob's, but at the same time
renounces by her own free choice to the
knowledge of position that she could have obtained on Bob's particle.
This renunciation is based
on the fact that we cannot predict the result of a measurement.
In~the example given above, if we want to measure the momentum of a
particle, since there is no way to predict either the momen\-tum
exchange or the displacement caused by it, we cannot know with
certainty where the particle interacted with the apparatus. 
The following quote summarizes Bohr's argument.

\begin{quote}\sf
In fact, the renunciation in each experimental arrangement of the
one or the other of two aspects of the description of physical phenomena,---the
combination of which characterizes the method of classical
physics, and which therefore in this sense may be considered as
\emph{complementary} to one another,---depends essentially on the
impossibility, in the field of quantum theory, of accurately
controlling the reaction of the object on the measuring
instruments, i.e.,~the transfer of momentum in case of position
measurements, and the displacement in case of momentum
measurements.
\end{quote}

The problem, of course, is that EPR had no issue with comple\-men\-tarity!
They did not claim in their paper that it would be possible to
\emph{know} simultaneously the position and
momentum of a particle with arbitrary precision.
Indeed, had they not agreed with this manifestation of
Heisenberg's uncertainty relations, they would surely have claimed
that QM is outright \emph{incorrect}, not merely incomplete!
Their disagreement with the view of QM is summarized in the following
quote from their paper:

\begin{quote}\sf
The usual conclusion from this\,%
\footnote{\,Measuring the coordinate of a particle alters its state.}
in quantum mechanics is that
\emph{when the momentum of a particle is known, its coordinate
has no physical reality}.
\end{quote}

\noindent
So, we see that, in the views of EPR, the incompleteness of QM has to do with the fact
that it does not allow simultaneous \emph{reality} for two non-commuting
observables. Whether or not these two realities could be learned
simultaneously is irrelevant to their argument.
The fact that QM allows us to predict
either the momentum or the position of a particle without having interacted with it,
mutually exclusive as these
measurements might be, seems to mean, according to
Definition~\ref{def:reality}, that both of these quantities have
physical reality. Therefore QM appears to be incomplete, according to their
criterion of Definition~\ref{def:completeness}.

%%%%%%%%%%%%%%%%%%%%%%%%%%%%%%%%%%%%%%%%%%%%%%%%%%%%%%%%%%%%%%%%%%%%%%%%%%%%%%%%%%%%%%%%%%

\section{Conclusion}\label{sec:conclusion}

We exhibited problems with the EPR argument,
when seen through the eyes of mathematical logic, without
any reference to Bell's inequality.
We~also questioned the narrow-minded view of ``states'' taken in
that paper, although that view is not surprising from advocates
of ``elements of reality''.
Speaking of which, we should like to add to
the criticisms laboured in our paper that the notion
of ``elements of reality'' is left rather vague in the EPR paper.
Nevertheless, we came down even harder on Bohr's response which,
in our opinion, had entirely failed to address the main issue raised by~EPR.

In retrospect, our criticism of the EPR paper notwithstanding,
we admit without shame that we would have been more likely to side with them
than with Bohr\ldots\ until the bolt gave way to thunder under Bell's pen!

%%%%%%%%%%%%%%%%%%%%%%%%%%%%%%%%%%%%%%%%%%%%%%%%%%%%%%%%%%%%%%%%%%%%%%%%%%%%%%%%%%%%%%%%%%

\section*{Acknowledgements}

The authors are grateful to Jeffrey Bub, Hilary Carteret, Christopher Fuchs
and N.~David Mermin for their comments and for helping us interpret
both the EPR paper and Bohr's response.
G.\,B. is supported in part by the Natural
Sciences and Engineering Research Council of Canada,
the Canada Research Chair programme and
the Canadian Institute for Advanced Research.

%%%%%%%%%%%%%%%%%%%%%%%%%%%%%%%%%%%%%%%%%%%%%%%%%%%%%%%%%%%%%%%%%%%%%%%%%%%%%%%%%%%%%%%%%%

\end{document}